\documentclass[12pt,preprint]{aastex}

\makeatletter

\newcommand{\Rmnum}[1]{\expandafter\@slowromancap\romannumeral#1@}
\makeatother
\usepackage{amsmath,amssymb}

\slugcomment{submission to ApJ Letters}
\shorttitle{Simulating solar prominences}
\shortauthors{Xia et al.}

\begin{document}
\title{Simulating the in situ condensation process of solar prominences}

\author{C. Xia\altaffilmark{1}, R. Keppens\altaffilmark{1,2}, 
  P. Antolin\altaffilmark{3}, O. Porth\altaffilmark{4}}

\altaffiltext{1}{Centre for mathematical Plasma Astrophysics, Department of 
Mathematics, KU Leuven, Celestijnenlaan 200B, 3001 Leuven, Belgium}
\altaffiltext{2}{School of Astronomy and Space Science, Nanjing University, Nanjing 210093, China}
\altaffiltext{3}{National Astronomical Observatory of Japan, 2-21-1 Osawa, 
Mitaka, Tokyo 181-8588, Japan}
\altaffiltext{4}{Department of Applied Mathematics, The University of Leeds, Leeds, LS2 9JT, UK}

\begin{abstract}
Prominences in the solar corona are hundredfold cooler and denser than 
their surroundings, with a total mass of $10^{13}$ up to $10^{15}$ g. 
Here we report on the first comprehensive simulations of three-dimensional, 
thermally and gravitationally stratified magnetic flux ropes, where in situ 
condensation to a prominence happens due to radiative losses. After a gradual 
thermodynamic adjustment, we witness a phase where runaway cooling happens
 while counter-streaming shearing flows drain off mass along helical field 
lines. After this drainage, a prominence-like condensation resides in 
concave upward field regions, and this prominence retains its overall 
characteristics for more than two hours. While condensing, the prominence 
establishes a prominence-corona transition region, where magnetic field-aligned
 thermal conduction is operative during the runaway cooling. The prominence 
structure represents a force-balanced state in a helical flux rope. The simulated 
condensation demonstrates a right-bearing barb, as a remnant of the drainage. 
Synthetic images at extreme ultraviolet wavelengths follow the onset of the 
condensation, and confirm the appearance of horns and a three-part structure 
for the stable prominence state, as often seen in 
erupting prominences. This naturally explains recent Solar Dynamics 
Observatory views with the Atmospheric Imaging Assembly on 
prominences in coronal cavities demonstrating horns.
\end{abstract}


\section{INTRODUCTION}\label{intro}
Solar observations provide detailed views on prominences, which consist of 
cool, dense material suspended in the corona in a central axial sheet-like 
filament spine, made of many threads and connected to lower altitudes by 
means of barbs~\citep{parenti14}. 
On the northern (southern) solar hemisphere, one encounters 
mostly dextral (sinistral) flux ropes with anticlockwise (clockwise) winding 
along the axis, wherein barbs preferentially form in a right-bearing 
(left-bearing) fashion \citep{Mart98,Chen14}. Prominences also show rich 
internal dynamics throughout, with counter-streaming plasma flows hinting at 
field connections down to the photosphere~\citep{zirkernat98}.
Recent Solar Dynamics Observatory 
\citep{sdo12} (SDO) observations using the Atmospheric Imaging Assembly 
\citep{aia12} (AIA) provided further evidence for flux rope (FR) topologies 
underpinning coronal cavities~\citep{forland13,gibson13}, the dark and 
low-density coronal tunnels surrounding the prominence proper. When viewed 
along the filament spine at the solar limb, extreme ultraviolet (EUV) 
observations detect prominence horns emanating into the 
cavity~\citep{gibson13}. All these aspects point to intrinsically 
three-dimensional condensation processes during 
formation, a phase only recently imaged in multiple EUV 
channels~\citep{LiuW12, Berger12}. 

Models that corroborate the interpretation of in situ 
condensation as a result of runaway cooling through thermal 
instability~\citep{Park53,Fiel65} have mainly been 
one-dimensional (1D) simulations~\citep{antiochos99,karpen01,xia11,luna12,
Schm13}. The 1D approach has been combined with rigid three-dimensional 
(3D) field lines from magnetohydrodynamic (MHD) simulations 
ignoring gravity and thermodynamics~\citep{vore05}, or from 
previous isothermal FR scenarios~\citep{fan05}. Although 
this combination gives hints of how projection effects matter within 
3D topologies~\citep{luna12,Schm13}, true 3D modeling is 
essential to understand the magnetic 
and thermal structure of prominences and their relation with surrounding 
coronal cavities, as is demonstrated and presented here. 

\section{SIMULATION STRATEGY}\label{meth}
We start with a FR structure already obeying 
macroscopic force balance between gravity, pressure gradients 
and the Lorentz force, and an overlying magnetic arcade that has a left-skewed 
orientation with respect to the FR axis, typical for dextral FR. 
This configuration was generated in an isothermal MHD 
simulation~\citep{xia14} at finite plasma beta, by subjecting a linear 
force-free arcade to vortical footpoint motions that alter the arcade 
magnetic shear, and then to converging footpoint motions towards the polarity 
inversion line (PIL). This boundary driven evolution established the 
characteristic sigmoidal or S-shaped FR. This gravitationally stratified, 
stable FR has an elliptical cross-sectional shape, and enough (order 
$10^{14}$ g) hot (1 MK) plasma to form a small prominence in the upwardly 
concave parts of the magnetic configuration. Based on this configuration, we 
start an MHD simulation using MPI-AMRVAC~\citep{amrvac12,porth14}.
The MHD equations have their usual form~\citep{xia12,xia14,
Keppens14}, including a total energy evolution with purely field-aligned 
thermal conduction of the form $\nabla\cdot\left(\kappa T^{2.5} \hat{\mathbf{e}}_B 
\hat{\mathbf{e}}_B \cdot \nabla T\right)$ (using the 
coefficient $\kappa=10^{-6}$ erg s$^{-1}$ cm$^{-1}$ K$^{-3.5}$), and tabulated 
losses through $Q\propto n_H^2 \Lambda(T)$ scaling with hydrogen number density 
squared, and a cooling table from~\cite{colganetal08}. In this work, the 
coronal heating term $H$ is parametrized as $H=C (B/B_0)^{0.5}\exp(-z/\lambda)$ with
 $C=2.2\times10^{-7}$ erg cm$^{-3}$ s$^{-1}$, $B_0=2$ G, $\lambda=120$ Mm, and $B$ 
denoting the local instantaneous field strength. We employ a shock-capturing 
scheme combining an HLL flux evaluation with third-order limited 
reconstruction~\citep{cada}, in a three-step Runge Kutta time 
marching~\citep{xia14,kepjcam14}. Thermal conduction is solved by 
a source-split strategy, using explicit sub-cycling within each time step.

As we start from an isothermal MHD configuration, we first modify the 
thermodynamics in the initial condition, to include a 
chromospheric layer. This modification affects a bottom layer from 3 Mm to 7 Mm, 
where the temperature is replaced by a hyperbolic tangent profile which connects 
a 10,000 K chromosphere to the 1 MK corona with a transition region at 6 Mm. The 
density in the bottom layer is then recalculated assuming hydrostatic 
equilibrium with bottom number density $10^{13}$ cm$^{-3}$. The simulation 
extends in $-120<x<120$ Mm, $-90<y<90$ Mm, and height $3<z<123$ 
Mm, with an effective mesh of $512\times384\times256$, using three grid 
levels. We impose zero velocity and extrapolate magnetic field with zero 
gradient (ensuring vanishing divergence) at all boundaries. We use  
zero-gradient extrapolation for density and pressure on sides, fixed 
gravitationally stratified density and pressure at bottom, and extrapolate
density and pressure at top via zero-gradient temperature extrapolation
assuming hydrostatic equilibrium.

\section{IN-SITU PROMINENCE FORMATION}\label{resul}
\begin{figure}
\includegraphics[width=5.8in]{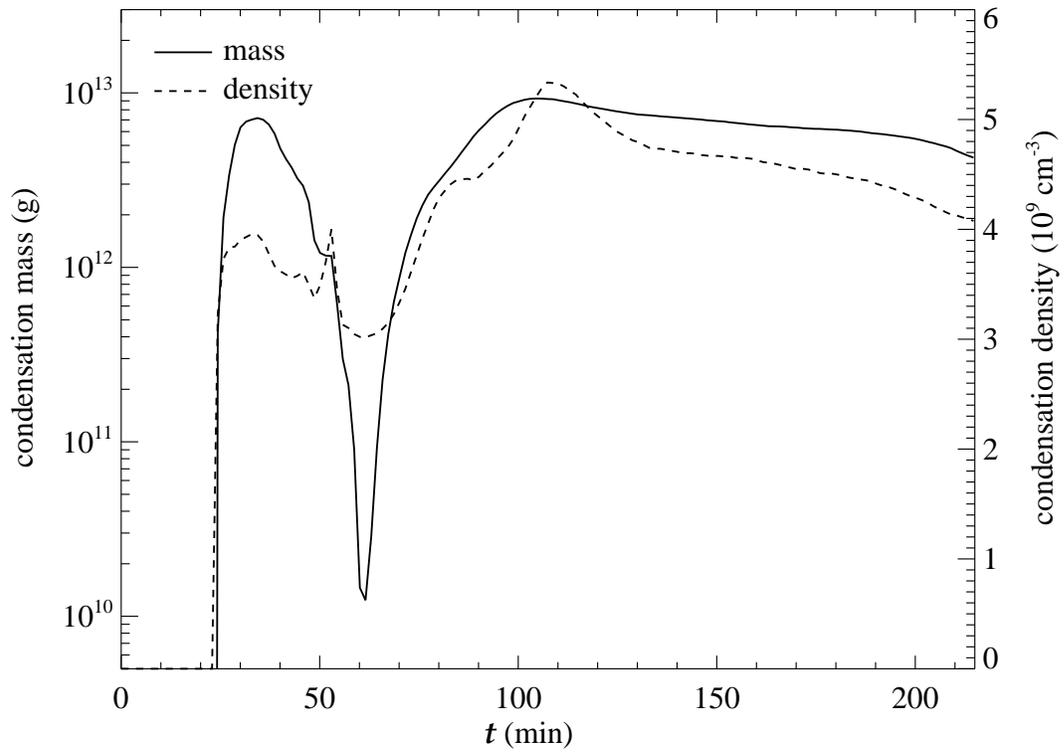}
\caption{Evolution of total mass (solid) and average number density (dashed) of the prominence.
}
\label{fmass}
\end{figure}

\begin{figure}
\includegraphics[width=5.8in]{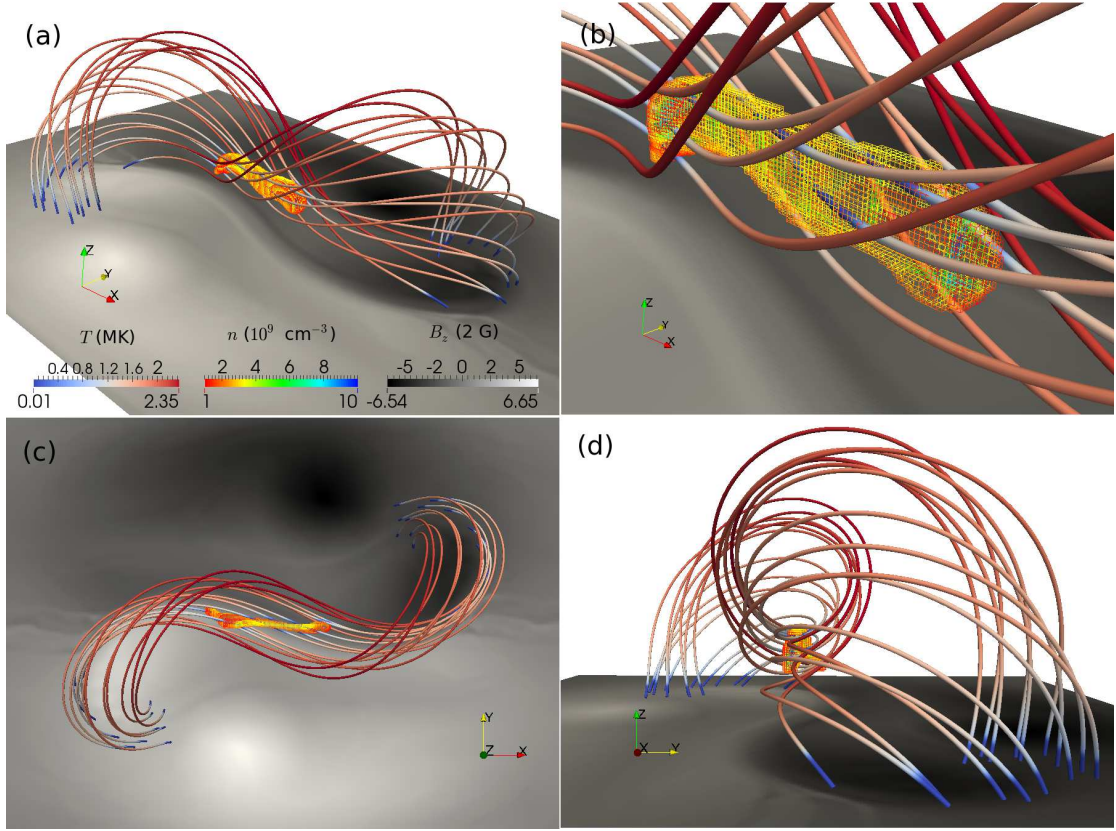}
\caption{The FR with embedded prominence at $t=150$ minutes. Shown are 
field lines colored by temperature in blue-red, the
 prominence colored by density in rainbow, and the bottom magnetogram 
in grey. Panels (a),(c), and (d) are side, top, and axial views, respectively. Panel (b) zooms into (a).
View (c) shows the filament spine within the S-shaped FR, with the barb near its left end. An animation is provided.
}
\label{fend}
\end{figure}

In the first 20 minutes, the combination of anisotropic thermal conduction, 
radiative losses, and parametrized heating cause readjustments from the 
initial state. The transition region remains situated at about 6 Mm height.
In a base layer immediately above the newly realized transition region, 
radiative loss exceeds coronal heating and the resulting decrease of gas 
pressure breaks force balance. This leads to some coronal plasma 
sliding down along arched field lines in the arcade and along helical field 
lines in the FR and adjusting to chromosphere conditions. Due to this weight 
loss, the FR rises slowly to a new 
balanced position with its axis now at 43 Mm height, 2 Mm higher than in 
the initial state. During previous FR formation 
process~\citep{xia14}, the formed helical field lines bring up high density 
plasma and compress plasma inside with Lorentz force toward the FR 
axis. Therefore the FR density became enhanced.
Although the coronal heating is tuned to approximately maintain the 
initial 1 MK temperature, thermal equilibrium does not exist in the FR 
initially where the local high density causes the temperature to 
decrease due to stronger radiative cooling than coronal heating.
At $t=21$ minutes, a central region 
at a height of 28 Mm reaches temperatures as low as 20,000 K, below which 
the optically thin radiative losses quickly diminish. To quantify this 
in situ condensation, Figure~\ref{fmass} shows the evolution of total mass 
and average number density of condensed plasma in the corona, i.e. denser 
than $3\times10^9$ cm$^{-3}$ and found above 9 Mm. This figure shows 
two phases: a first dynamic phase up to about 100 minutes, followed by a 
more stable evolution up to 215 minutes. Starting with the 
dynamic phase, the localized cooling creates a low gas pressure well, which 
sucks in ambient plasma along field lines, up to a point where its density 
increases up to 25 times the coronal value. This in situ plasma condensation 
attracts coronal plasma to the middle, while simultaneously 
counteracted by downflows that drain off matter to two feet of the FR 
 on both sides of the PIL. The S-shaped topology of the FR enforces this 
drainage to happen in opposite directions on each side of the FR central 
axis, so a spontaneous counter-streaming shearing flow develops 
inside the FR. These counter-streaming flows stretch and attenuate 
the ongoing condensation which partially streams 
from the center to both ends of the FR along helical magnetic field 
lines. While the condensation density can thereby go down to $2\times10^9$ 
cm$^{-3}$, its temperature remains below 30,000 K at all times. The maximal 
velocity of drained condensation fragments in this dynamic phase reaches up
 to 100 km s$^{-1}$, with typical values of 25 km s$^{-1}$ throughout the 
FR. This can be compared with the few 10 km s$^{-1}$ originally 
reported~\citep{zirkernat98} for counter-streaming flows in prominences. 
This counter-streaming leads to the sharp drop in mass and density seen in 
Figure~\ref{fmass} at around $t=60$ minutes, but gradually slows down, 
leaving behind a condensation fragment in the concave upward field region 
which reaches a maximal density of $5.3\times10^9$ cm$^{-3}$ at $t=107$ 
minutes, remaining at a total mass between $10^{13}$ g down to 
$4\times10^{12}$ g throughout. This condensation is analogous to a 
prominence, which stays approximately motionless for several hours. 
We visualize a stable state at $t=150$ minutes in Figure~\ref{fend} 
(an animated view from varying perspectives is additionally provided as a movie). 
We show four views, each time containing selected helical field 
lines colored by temperature, the prominence in a wireframe mesh view as 
colored by density, and the bottom magnetogram. Panels (a) \& (b) show side 
views with a zoom to the prominence body in Panel (b). Top and axial views 
are shown in Panels (c) \& (d), respectively. The prominence is visualized by 
a wireframe connecting all cells where the density exceeds $3\times10^9$ 
cm$^{-3}$. The prominence has a roughly slab-like shape above the PIL, with 
the bottom edge showing more curved variation than the top. Near the left 
end of the prominence in Panel (c), a branch protrudes to the right side of 
the prominence axis, leaning towards lower altitudes. This protrusion is 
reminiscent of prominence barbs, although we emphasize that no parasitic 
polarity in the bottom magnetogram is present. The barb develops at 
the site where most prominence mass first regathers and bends down local 
field lines, following the counter-streaming drainage to both feet, where 
simulation asymmetries cause this to happen off-center. This differs 
from the interpretation where barbs or feet 
extend from the prominence body down to parasitic polarity patches in the 
photosphere, explored in linear force-free models by~\citet{aulanier98}. 
The barb in our simulation obeys the right-bearing character, as its 
original flow pattern followed the helical field lines. The total size of the 
simulated prominence is about 46.4 Mm long, 13.1 Mm tall, and 4.8 Mm thick 
and its top reaches 26 Mm height. Its density ranges from $3\times10^9$ 
cm$^{-3}$ to $1\times10^{10}$ cm$^{-3}$ with on average $4.7\times10^9$ 
cm$^{-3}$, while coronal plasma at the same altitudes has an average density of
 $1.6\times10^8$ cm$^{-3}$. Since the magnetic field threading through the 
prominence is pointing from negative regions to positive in the underlying 
magnetogram, it is an inverse-polarity prominence. The field strength in the 
prominence increases slightly with height from 7.5 G at the bottom to 8.8 G at 
the top with an average value of 8.2 G. The angle between the prominence axis 
and the magnetic field vector in the horizontal plane is around 18$^\circ$. At 
the prominence lateral boundaries, the field makes an angle to the horizontal 
plane of around 29$^\circ$, consistent with observations~\citep{Bommi94}. 

\section{SYNTHETIC EUV VIEWS}\label{aia}
We synthesize quantities comparable with remote sensing observations 
following~\citet{Mok05}. The flux of optically 
thin emission measured by an imaging instrument in a certain wavelength band 
$i$ is treated as a line-of-sight (LOS) integral through the emitting plasma, 
\begin{equation}
D_i=\int n_{\rm e}^2 G_i(n_{\rm e}, T_{\rm e})\,dl~~~~~[{\rm DN~s}^{-1}],
\end{equation}
where $l$ is distance along the line of sight. The pixel
 response $D_i$ is in Data Number (DN) per second.
The instrumental response $G_i(n_{\rm e}, T_{\rm e})$ is function 
of electron density $n_{\rm e}$ and temperature $T_{\rm e}$, and takes into 
account atomic physics and instrument properties in band $i$. 
We use the CHIANTI 7 catalogue \citep{Landi12} and AIA routines \citep{Boe12} 
in SolarSoft. The LOS integral is 
evaluated by interpolation-based ray-tracing with a uniform grid of rays 
passing through the block-adaptive octree grid. We make synthetic images along the
$x$-axis direction in four EUV wavelength bands 304, 171, 193, and 211 \AA~of 
the SDO/AIA instrument, which sample temperatures from 0.08, 0.8, 1.5 
 and 1.8 MK, respectively. We describe the onset of the in situ process as
 seen in EUV images, and then discuss synthetic views for the prominence end state.

\begin{figure}
\includegraphics[width=6.in]{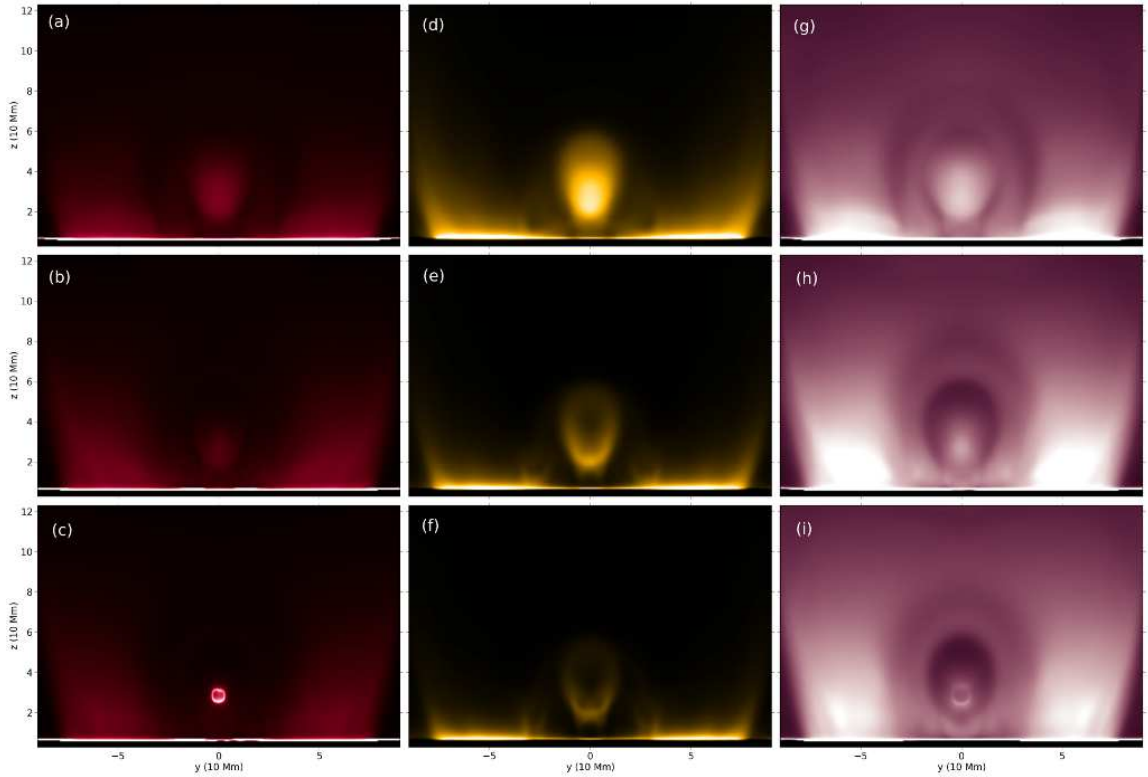}
\caption{
Synthetic SDO/AIA views showing the condensation
process. We view along the $x$-axis at time 11.4, 17.2, and 21.5 minutes 
from top to bottom, respectively. Approximate wavelength and peak 
temperature sensitivity are from left to right: 304 \AA~(0.08 MK), 171 
\AA~(0.8 MK), and 211 \AA~(1.8 MK). 
}
\label{fevo}
\end{figure}

\begin{figure}
\includegraphics[width=6.in]{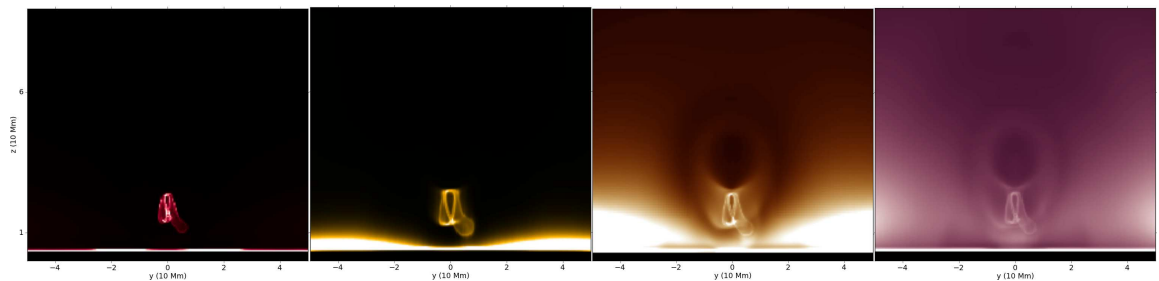}
\caption{Synthetic SDO/AIA views at $t=150$ minutes. The wavelength and peak 
temperature are from left to right: 304 \AA~(0.08 MK), 171 
\AA~(0.8 MK), 193 \AA~(1.5 MK), and 211 \AA~(1.8 MK). In the latter 
two wavelengths we see prominence horns 
and a central dark cavity.
}
\label{fsyn}
\end{figure}

\begin{figure}
\includegraphics[width=5.8in]{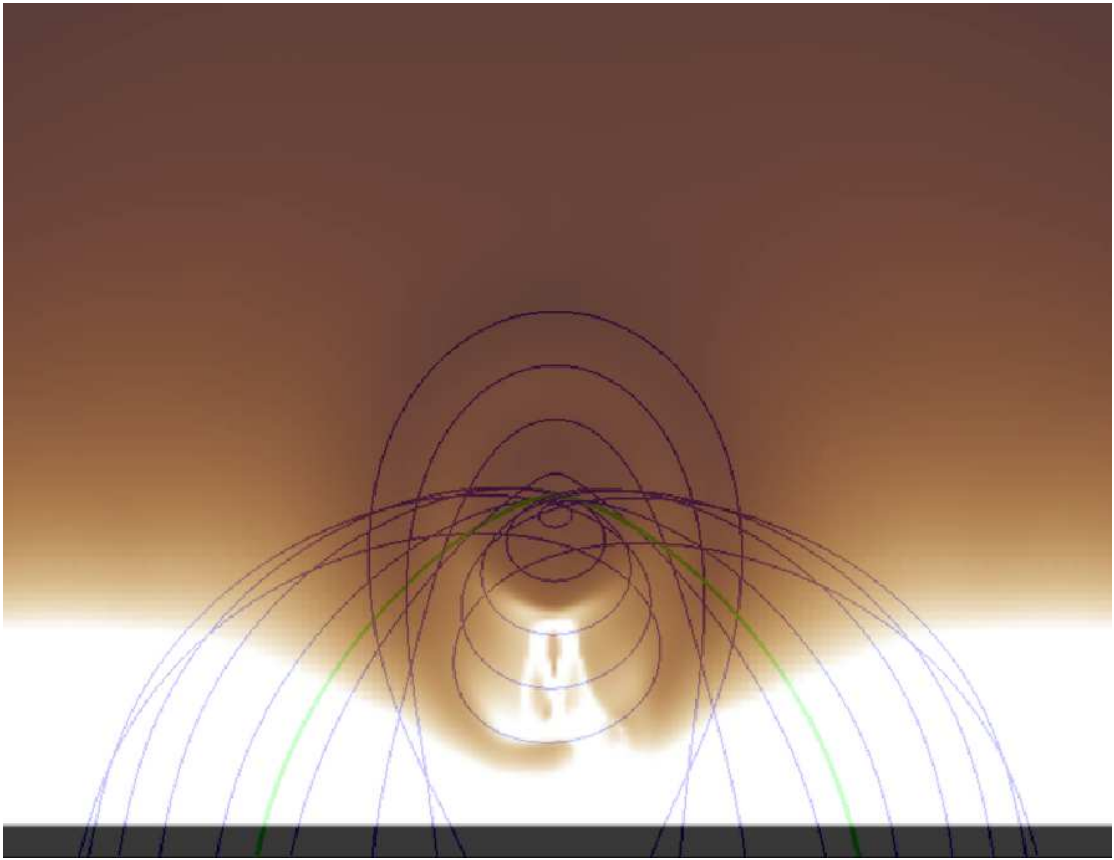}
\caption{Magnetic field lines overlaying the synthetic SDO/AIA
 193 \AA~view, at $t=150$ minutes. The axis of the FR is the thick green line.
}
\label{fcav}
\end{figure}

In Figure~\ref{fevo}, representative synthetic views along the $x$-axis, 
taken at times 11.4, 17.2, and 21.5 minutes from the top row to the bottom 
respectively, reveal the dynamic process of plasma condensation as seen in 304, 171, and
211 \AA~EUV bands. The 304 \AA~channel roughly corresponds to 0.08 MK, but has another emission 
contribution from the Si \Rmnum{11} 303.33 \AA~line formed at a coronal 
temperature of 1.6 MK. Although its response is four times weaker than the main 
He \Rmnum{2} lines of this channel, its emission contribution dominates in the
off-limb corona in this 304 \AA~channel \citep{ODwy10}. Therefore, the million Kelvin, 
relatively dense coronal plasma trapped in the lower part of the FR 
originally produces a bright dispersive cloud in this 304 \AA~channel as shown in panel (a).
At the corresponding time of 11.4 minutes, this cloud plasma temperature ranges from 0.7 MK to 0.8 MK, 
and is most prominent in the 171 \AA~channel but also visible at 211
\AA. The temperature response in this 211 \AA~channel has a weak
wide-spread contribution from 0.16 MK to 1 MK, although it peaks at 1.8 MK. Due to 
the initial gradual thermodynamic adjustment, the center of the cloud 
progressively cools from 1 MK to 0.02 MK during the first 20 minutes, while the 
density at the center increases by 67 \%. The resulting pressure change 
generates flows into the cooling core with speed up to 74 km s$^{-1}$.
Because hot channel emission decreases as the temperature drops, 
the bright cloud is seen to fade by all these EUV views from the top row to the middle
row. As the counter-streaming dynamics removes mass from the cloud to both ends of 
the FR as mentioned before, the outer layer of the cloud gets attenuated 
and further darkens the cloud. In panel (c), a bright core forms in the
FR at 28 Mm height where the temperature at the center of the cloud
drops as low as 0.02 MK. This cool core marks the site where thermal 
instability sets in, which subsequently dramatically increases in density 
while growing in spatial extent to a large-scale prominence. At this early 
onset phase, this cool core is at first dark in the other two hot channels
(see panel (f) and (i)), while later on it is seen with bright edges in 
Figs.~\ref{fsyn}-\ref{fcav}.
The bright ring around this cool core in the 304 \AA~channel is due
to the optically thin treatment where we cumulatively integrate strong
emission along the prominence-corona transition regions (PCTR). In reality,
prominences have more uniform luminance in the 304 \AA~channel, because 
prominence plasma is optically thick for the EUV line He \Rmnum{2} 304 \AA~and 
most contribution is due to scattering of this EUV emission from the sun 
\citep{Labrosse12,Labrosse07}.
In 211 \AA~channel views, we witness the formation of a dark cavity simultaneously
with the prominence formation. The cavity has an elliptic shape and encompasses the
 forming prominence. In this early phase, the formation of the cavity is 
primarily due to cooling of the dense FR region, since the density 
at the same altitude is comparable with or even larger than the surrounding hot
 arcade. Later on, as the mass drainage of the FR continues, the cavity 
temperature recovers to values above a million Kelvin and the cavity density drops
below the value of the surrounding coronal arcade. The dark bottom layers of the
chromosphere shown no emission in EUV lines due to their low temperature.

The prominence loaded FR in Figure~\ref{fend} maintains its shape 
at approximately constant total mass for about two hours. We take the snapshot 
at $t=150$ minutes to analyze this stable state with synthetic views in all four EUV bands, 
shown in Figure~\ref{fsyn}. From these, one detects that the PCTR shows stronger 
emission than internal prominence regions, and prominently appears in outline in 
all bands. This is because the temperature variation in the PCTR yields (cumulative) emission
contributions to all four EUV bands in the optically thin approximation we 
adopted. The lower part of the prominence is seen to tilt to the left 
near its front end and to the right when close to the back end. These front 
and back PCTR outlines overlap and give a strong emission spot near the bottom. 
The right-protruding tail or barb extends to lower altitude, as typical for 
prominence barbs. The prominence in 304 \AA~is slightly thinner than in the 
hotter bands, since cooler cores are enveloped by hotter layers. In 193 \AA, 
the coronal dark cavity is most noticeable. In 193 \AA~and 211 \AA~channels, 
one can detect the horn like structures that extend from the top of the 
prominence to the upper cavity. The extension of these horns connect in the 
top region of the cavity, which forms a closed ellipse dividing the cavity into 
two parts, namely an inner elliptical dark region and an outer dark ring. The
density distribution inside the cavity is 20 \% to 30 \% lower than in the
surrounding arcade at the same height, while the temperature is slightly
higher, about 2 MK. These density and temperature values of the cavity are
consistent with observations \citep{gibson13,Fuller09}. 
In fact, we can quantify the field topology precisely, and therefore combine a 
synthetic view in 193 \AA, with projected field line views in Figure~\ref{fcav}. 
We saturated the view in the lower corona, to show the cavity and horns with 
better contrast. The axis of the FR is the thicker green
field line in Figure~\ref{fcav}. It is then seen that there are two
kinds of field lines in the FR threading through the dark cavity. 
Arched field lines, with their central points above the axis of the FR, 
have been twisted but have no concave upward parts to collect and support 
prominence plasma. Concave upward 
helical field lines, with their central points below the FR axis, have 
progressively larger concave parts as they reach lower altitudes. The outer dark 
ring is threaded by twisted arched loops overlying the FR. For all these 
field lines, their paired footpoints are closer to the PIL and each other when 
their central points are higher. 
The horns are actually LOS emission from prominence-loaded helical field 
lines that maintain denser coronal plasma than prominence-free field lines of the 
cavity. During the cavity-prominence formation, density 
depletion occurs not only on prominence-loaded field lines threading  
cavity and prominence where in situ condensation happens~\citep{Berger12}, but also on  
prominence-free field lines due to mass drainage into the chromosphere. 
The magnetic structure changes slowly and 
smoothly from the horns to the central cavity on top of the prominence, 
although the thermal structure changes significantly. 

\section{CONCLUSIONS AND DISCUSSION}\label{conclusion}
We demonstrated in situ condensation occuring in a dextral FR
configuration, leading to a macroscopic prominence. Using synthetic SDO/AIA 
views, both the onset as well as the final prominence appearance is analyzed. 
The establishment of a coronal cavity surrounding the prominence is discussed 
in relation to the field topology and the evolving thermodynamics. The end 
state prominence is relatively low in total mass, but displays many 
characteristics in line with recent observations, including the distinctive 
horns as seen at EUV wavelengths. Before settling into a stable prominence
 configuration, counterstreaming flows on both FR halves develop. 
As a remnant of this flow, a barb persists on one side of the prominence spine. 
We plan to explore variations of the imposed heating $H$ (e.g. using 
impulsive heating as in 1D models~\cite{Karp08}) and of the initial FR
dimension and topology, following our strategy to start from an isothermally 
formed FR. Future work will need to include thermal condensation in 
self-consistent simulations of FR formation, and handle the optically 
thick conditions in the prominence and PCTR.

\acknowledgments
This research was supported by projects GOA/2015-014 (2014-2018 KU Leuven), 
FWO Pegasus, and the Interuniversity Attraction Poles Programme 
by the Belgian Science Policy Office (IAP P7/08 CHARM). The 
simulations used the VSC (flemish supercomputer center) funded by Hercules 
foundation and Flemish government.

\bibliographystyle{apj}
%\bibliography{apjsubmission}

\begin{thebibliography}{39}
\expandafter\ifx\csname natexlab\endcsname\relax\def\natexlab#1{#1}\fi

\bibitem[{{Antiochos} {et~al.}(1999){Antiochos}, {MacNeice}, {Spicer}, \&
  {Klimchuk}}]{antiochos99}
{Antiochos}, S.~K., {MacNeice}, P.~J., {Spicer}, D.~S., \& {Klimchuk}, J.~A.
  1999, \apj, 512, 985

\bibitem[{{Aulanier} \& {Demoulin}(1998)}]{aulanier98}
{Aulanier}, G. \& {Demoulin}, P. 1998, \aap, 329, 1125

\bibitem[{{Berger} {et~al.}(2012){Berger}, {Liu}, \& {Low}}]{Berger12}
{Berger}, T.~E., {Liu}, W., \& {Low}, B.~C. 2012, \apj, 758, L37

\bibitem[{{Boerner} {et~al.}(2012){Boerner}, {Edwards}, {Lemen}, {Rausch},
  {Schrijver}, {Shine}, {Shing}, {Stern}, {Tarbell}, {Title}, {Wolfson},
  {Soufli}, {Spiller}, {Gullikson}, {McKenzie}, {Windt}, {Golub}, {Podgorski},
  {Testa}, \& {Weber}}]{Boe12}
{Boerner}, P., {Edwards}, C., {Lemen}, J., {Rausch}, A., {Schrijver}, C.,
  {Shine}, R., {Shing}, L., {Stern}, R., {Tarbell}, T., {Title}, A., {Wolfson},
  C.~J., {Soufli}, R., {Spiller}, E., {Gullikson}, E., {McKenzie}, D., {Windt},
  D., {Golub}, L., {Podgorski}, W., {Testa}, P., \& {Weber}, M. 2012, \solphys,
  275, 41

\bibitem[{{Bommier} {et~al.}(1994){Bommier}, {Landi Degl'Innocenti}, {Leroy},
  \& {Sahal-Brechot}}]{Bommi94}
{Bommier}, V., {Landi Degl'Innocenti}, E., {Leroy}, J.-L., \& {Sahal-Brechot},
  S. 1994, \solphys, 154, 231

\bibitem[{{Chen} {et~al.}(2014){Chen}, {Harra}, \& {Fang}}]{Chen14}
{Chen}, P.~F., {Harra}, L.~K., \& {Fang}, C. 2014, \apj, 784, 50

\bibitem[{{Colgan} {et~al.}(2008){Colgan}, {Abdallah}, {Sherril}, {Foster},
  {Fontes}, \& {Feldman}}]{colganetal08}
{Colgan}, J., {Abdallah}, J.~J., {Sherril}, M., {Foster}, M., {Fontes}, C.~J.,
  \& {Feldman}, U. 2008, \apj, 689, 585

\bibitem[{{DeVore} {et~al.}(2005){DeVore}, {Antiochos}, \& {Aulanier}}]{vore05}
{DeVore}, C.~R., {Antiochos}, S.~K., \& {Aulanier}, G. 2005, \apj, 629, 1122

\bibitem[{{Fan}(2005)}]{fan05}
{Fan}, Y. 2005, \apj, 630, 543

\bibitem[{{Field}(1965)}]{Fiel65}
{Field}, G.~B. 1965, \apj, 142, 531

\bibitem[{{Forland} {et~al.}(2013){Forland}, {Gibson}, {Dove}, {Rachmeler}, \&
  {Fan}}]{forland13}
{Forland}, B.~C., {Gibson}, S.~E., {Dove}, J.~B., {Rachmeler}, L.~A., \& {Fan},
  Y. 2013, \solphys, 288, 603

\bibitem[{{Fuller} \& {Gibson}(2009)}]{Fuller09}
{Fuller}, J. \& {Gibson}, S.~E. 2009, \apj, 700, 1205

\bibitem[{{Illing} \& {Hundhausen}(1985)}]{illing95}
{Illing}, R.~M.~E. \& {Hundhausen}, A.~J. 1985, \jgr, 90, 275

\bibitem[{{Karpen} {et~al.}(2001){Karpen}, {Antiochos}, {Hohensee}, {Klimchuk},
  \& {MacNeice}}]{karpen01}
{Karpen}, J.~T., {Antiochos}, S.~K., {Hohensee}, M., {Klimchuk}, J.~A., \&
  {MacNeice}, P.~J. 2001, \apjl, 553, L85

\bibitem[{{Karpen} \& {Antiochos}(2008)}]{Karp08}
{Karpen}, J.~T. \& {Antiochos}, S.~K. 2008, \apj, 676, 658

\bibitem[{{Keppens} {et~al.}(2012){Keppens}, {Meliani}, {van Marle}, {Delmont},
  {Vlasis}, \& {van der Holst}}]{amrvac12}
{Keppens}, R., {Meliani}, Z., {van Marle}, A.~J., {Delmont}, P., {Vlasis}, A.,
  \& {van der Holst}, B. 2012, Journal of Computational Physics, 231, 718

\bibitem[{{Keppens} \& {Porth}(2014)}]{kepjcam14}
{Keppens}, R. \& {Porth}, O. 2014, Journal of Computational and Applied
  Mathematics, 266, 87

\bibitem[{{Keppens} \& {Xia}(2014)}]{Keppens14}
{Keppens}, R. \& {Xia}, C. 2014, \apj, 789, 22

\bibitem[{{Landi} {et~al.}(2012){Landi}, {Del Zanna}, {Young}, {Dere}, \&
  {Mason}}]{Landi12}
{Landi}, E., {Del Zanna}, G., {Young}, P.~R., {Dere}, K.~P., \& {Mason}, H.~E.
  2012, \apj, 744, 99

\bibitem[{{Lemen} {et~al.}(2012){Lemen}, {Title}, {Akin}, {Boerner}, {Chou},
  {Drake}, {Duncan}, {Edwards}, {Friedlaender}, {Heyman}, {Hurlburt}, {Katz},
  {Kushner}, {Levay}, {Lindgren}, {Mathur}, {McFeaters}, {Mitchell}, {Rehse},
  {Schrijver}, {Springer}, {Stern}, {Tarbell}, {Wuelser}, {Wolfson}, {Yanari},
  {Bookbinder}, {Cheimets}, {Caldwell}, {Deluca}, {Gates}, {Golub}, {Park},
  {Podgorski}, {Bush}, {Scherrer}, {Gummin}, {Smith}, {Auker}, {Jerram},
  {Pool}, {Soufli}, {Windt}, {Beardsley}, {Clapp}, {Lang}, \&
  {Waltham}}]{aia12}
{Lemen}, J.~R., et~al. 2012, \solphys, 275, 17

\bibitem[{{Labrosse} {et~al.}(2007){Labrosse}, {Gouttebroze}, 
\& {Vial}}]{Labrosse07}
{Labrosse}, N., {Gouttebroze}, P., \& {Vial}, J.-C. 2007, \aap, 463, 1171 

\bibitem[{{Labrosse} \& {McGlinchey}(2012)}]{Labrosse12} 
{Labrosse}, N., \& {McGlinchey}, K. 2012, \aap, 537, A100 

\bibitem[{{Liu} {et~al.}(2012){Liu}, {Berger}, \& {Low}}]{LiuW12}
{Liu}, W., {Berger}, T.~E., \& {Low}, B.~C. 2012, \apjl, 745, L21

\bibitem[{{Luna} {et~al.}(2012){Luna}, {Karpen}, \& {DeVore}}]{luna12}
{Luna}, M., {Karpen}, J.~T., \& {DeVore}, C.~R. 2012, \apj, 746, 30

\bibitem[{{Martin}(1998)}]{Mart98}
{Martin}, S.~F. 1998, \solphys, 182, 107

\bibitem[{{Mok} {et~al.}(2005){Mok}, {Miki{\'c}}, {Lionello}, \&
  {Linker}}]{Mok05}
{Mok}, Y., {Miki{\'c}}, Z., {Lionello}, R., \& {Linker}, J.~A. 2005, \apj, 621,
  1098

\bibitem[{{O'Dwyer} {et~al.}(2010){O'Dwyer}, {Del Zanna}, {Mason}, {Weber}, \&
  {Tripathi}}]{ODwy10}
{O'Dwyer}, B., {Del Zanna}, G., {Mason}, H.~E., {Weber}, M.~A., \& {Tripathi},
  D. 2010, \aap, 521, A21

\bibitem[{{Parenti}(2014)}]{parenti14}
{Parenti}, S. 2014, Living Reviews in Solar Physics, 11, 1

\bibitem[{{Parker}(1953)}]{Park53}
{Parker}, E.~N. 1953, \apj, 117, 431

\bibitem[{{Pesnell} {et~al.}(2012){Pesnell}, {Thompson}, \&
  {Chamberlin}}]{sdo12}
{Pesnell}, W.~D., {Thompson}, B.~J., \& {Chamberlin}, P.~C. 2012, \solphys,
  275, 3

\bibitem[{{Porth} {et~al.}(2014){Porth}, {Xia}, {Hendrix}, {Moschou}, \&
  {Keppens}}]{porth14}
{Porth}, O., {Xia}, C., {Hendrix}, T., {Moschou}, S.~P., \& {Keppens}, R. 2014,
  Accepted by \apjs

\bibitem[{{Schmit} \& {Gibson}(2013)}]{gibson13}
{Schmit}, D.~J. \& {Gibson}, S. 2013, \apj, 770, 35

\bibitem[{{Schmit} {et~al.}(2013){Schmit}, {Gibson}, {Luna}, {Karpen}, \&
  {Innes}}]{Schm13}
{Schmit}, D.~J., {Gibson}, S., {Luna}, M., {Karpen}, J., \& {Innes}, D. 2013,
  \apj, 779, 156

\bibitem[{{{\v C}ada} \& {Torrilhon}(2009)}]{cada}
{{\v C}ada}, M. \& {Torrilhon}, M. 2009, Journal of Computational Physics, 228,
  4118

\bibitem[{{Xia} {et~al.}(2012){Xia}, {Chen}, \& {Keppens}}]{xia12}
{Xia}, C., {Chen}, P.~F., \& {Keppens}, R. 2012, \apj, 748, L26

\bibitem[{{Xia} {et~al.}(2011){Xia}, {Chen}, {Keppens}, \& {van Marle}}]{xia11}
{Xia}, C., {Chen}, P.~F., {Keppens}, R., \& {van Marle}, A.~J. 2011, \apj, 737,
  27

\bibitem[{{Xia} {et~al.}(2014){Xia}, {Keppens}, \& {Guo}}]{xia14}
{Xia}, C., {Keppens}, R., \& {Guo}, Y. 2014, \apj, 780, 130

\bibitem[{{Zirker} {et~al.}(1998){Zirker}, {Engvold}, \&
  {Martin}}]{zirkernat98}
{Zirker}, J.~B., {Engvold}, O., \& {Martin}, S.~F. 1998, Nature, 396, 440

\end{thebibliography}

\clearpage

\clearpage

\clearpage

\clearpage

\clearpage

\end{document}